\documentclass[journal]{IEEEtran}

\ifCLASSINFOpdf

\else

\fi
\usepackage{balance}
\usepackage{lipsum}       
\usepackage{changepage}   
\usepackage{url}
\usepackage{multirow,array}
\usepackage{cite}
\usepackage{graphicx}
\usepackage{hyperref}
\usepackage{caption}
 \usepackage{verbatim}
\usepackage{amsmath}
\usepackage{amssymb}

\usepackage{enumitem}
\usepackage{graphicx}
\usepackage{caption}
\usepackage{subcaption}
\usepackage[numbers,sort&compress]{natbib}
\makeatletter 

\newcommand{\Rmnum}[1]{\expandafter\@slowromancap\romannumeral #1@}
\makeatother
\begin{document}

\title{Spatial Image Steganography Based on Generative Adversarial Network}

\author{Jianhua Yang, Kai Liu,~\IEEEmembership{Student Member,~IEEE,}
        Xiangui Kang$^*$,~\IEEEmembership{Senior Member,~IEEE,}\\
        Edward K.Wong,~\IEEEmembership{Senior Member,~IEEE,}
        Yun-Qing Shi,~\IEEEmembership{Life Fellow,~IEEE }

\thanks{This work was supported by NSFC (Grant Nos. U1536204, 61772571,61702429), and the special funding for basic scientific research of Sun Yat-sen University (6177060230). (Corresponding author: Xiangui Kang.)

 J. Yang, K. Liu, X. Kang are with Guangdong Key Lab of Information Security, School of Data
and Computer Science, Sun Yat-Sen University, Guangzhou, China 510006, (e-mail:
isskxg@mail.sysu.edu.cn).

E. K. Wong is with Department of Computer Science and Engineering, New York University, Tandon School of Engineering, Brooklyn, NY 11201, (e-mail:ewong@nyu.edu).

Y. Shi is with Department of ECE, New Jersey Institute of Technology, Newark, NJ, USA 07102, (e-mail:shi@njit.edu).
 }
}

\maketitle

\begin{abstract}
With the recent development of deep learning on steganalysis, embedding secret information into digital images faces great challenges. In this paper, a secure steganography algorithm by using adversarial training is proposed. The architecture contain three component modules: a generator, an embedding simulator and a discriminator. A generator based on U-NET to translate a cover image into an embedding change probability is proposed. To fit the optimal embedding simulator and propagate the gradient, a function called Tanh-simulator is proposed. As for the discriminator, the selection-channel awareness (SCA) is incorporated to resist the SCA based steganalytic methods. Experimental results have shown that the proposed framework can increase the security performance dramatically over the recently reported method ASDL-GAN, while the training time is only 30\% of that used by ASDL-GAN. Furthermore, it also performs better than the hand-crafted steganographic algorithm S-UNIWARD.

\end{abstract}


\IEEEpeerreviewmaketitle

\section{Introduction}
\label{sec:introduction}
 Image steganography is a kind of technology to embed secret information into a cover image without drawing suspicion. With the development of steganalysis methods, it becomes a great challenge to design a secure stegnographic scheme. Because the efficient coding schemes \cite{STC} can embed messages close to the payload-distortion bound, the main task of image steganography is to minimize a well designed additive distortion function. In an adaptive steganography scheme, every pixel is assigned a cost to quantify the effect of making modification and the distortion is evaluated by summing up costs. Secret information is generally embedded in noisy regions or regions with texture, while smooth regions are avoided for data embedding, as done by HUGO \cite{HUGO}, WOW \cite{WOW}, HILL \cite{HILL}, S-UNIWARD \cite{UNIWARD} and MiPOD \cite {MiPOD}.

In recent years, convolutional neural networks (CNN) have become a dominant machine learning approach in image classification tasks with the improvements in computer hardware and network architecture \cite{ResNet,DSN}. Current researches have indicated that CNN also obtained considerable achievements in the field of steganalysis. Tan and Li \cite{tan2014stacked} used the stacked convolutional auto-encoder for steganalysis. Qian \textit{et~al} \cite{qian2015deep,qian2016} proposed a CNN structure equipped with Gaussian non-linear activation, and they showed that feature representations can be transferred from high embedding payload to low embedding payload. Xu \textit{et~al} \cite{Xustructdesign, Xuensemble} proposed a CNN structure (referred to as XuNet in this paper) that is able to achieve comparable performance to conventional spatial rich model (SRM) \cite{SRM}. The Tanh and ReLU have been used in shallow and deep layers respectively. Batch-normalization \cite{BN} was equipped to prevent the network from falling into local minima. Yang \textit{et~al} \cite{Yang2017} incorporated selection-channel awareness (SCA) into the CNN architecture. Ye \textit{et~al} \cite{Ni2017} proposed a structure that incorporates high-pass filters from SRM, the SCA also be incorporated in CNN architecture. In \cite{yang2017jpeg}, Yang \textit{et~al} proposed a deep learning architecture by improving the pre-processing layer and the feature reuse for JPEG steganalysis, experimental results shows that it can obtain state-of-the art performance for JPEG steganalysis. Although CNN has been successfully used for steganalysis, it is still in initial stage with regarding to applying it for steganography.

So far, the generative adversarial network (GAN) \cite{goodfellow2014generative} has been widely used for image generation \cite{DCGAN,generateimge1}. In \cite{tang2017automatic}, Tang \textit{et~al} proposed an automatic steganographic distortion learning framework with GAN (named as ASDL-GAN shortly). The probability of data embedding is learned via the adversarial training between the generator and the discriminator. The generator contains 25 groups, with every group starts with a convolutional layer, followed by batch normalization and a ReLU layer. The architecture of XuNet was employed as the discriminator. In order to fit the optimal embedding simulator \cite{Simulator} as well as propagate the gradient in back propagation, they proposed a ternary embedding simulator (TES) activation function. The reported experimental results showed that ASDL-GAN can learn steganographic distortions, but the performance is still inferior to the conventional steganographic scheme S-UNIWARD.

In this work, we propose a new GAN-based steganographic framework. Compared with the previous method ASDL-GAN \cite{tang2017automatic}, the main contributions of this paper are as follows.
\begin{enumerate}[label= (\arabic*)]

	\item An activation function called Tanh-simulator is proposed to solve the problem that optimal embedding simulator cannot propagate gradient. The TES sub-network of ASDL-GAN needs a long time to pre-train with even $10^{6}$ iterations, while the Tanh-simulator can be used directly with high fitting accuracy.
	\item A more compact generator based on U-NET \cite{UNET} has been proposed. This subnet can improves security performance and decreases training time dramatically.
	\item Considering adversarial training, we enhance the discriminator by incorporating SCA into the discriminator to improve the performance of resisting SCA based steganalytic schemes.
\end{enumerate}

The rest of the paper is organized as follows. The proposed architecture is described in Section \Rmnum{2}. Experimental results and analysis is shown in Section \Rmnum{3}. The practical application of the proposed architecture is shown in Section \Rmnum{4}. The conclusion and future works are presented in Section \Rmnum{5}.

\section{The Proposed architecture}
In this section, firstly, we present the overall architecture of the proposed method based on generative adversarial network (referred as UT-SCA-GAN), which incorporates the U-net based generator, the proposed Tanh-simulator function and the SCA based discriminator. Secondly, the definition of the loss functions is introduced. Then, the details of the generator and the proposed Tanh-simulator function are described. Finally, we present the design consideration of the discriminator.
\subsection {Overall Architecture}
\begin{figure*}[!htb]
	\centering
	\includegraphics[totalheight=3.5in, width=5.5in, origin=c]{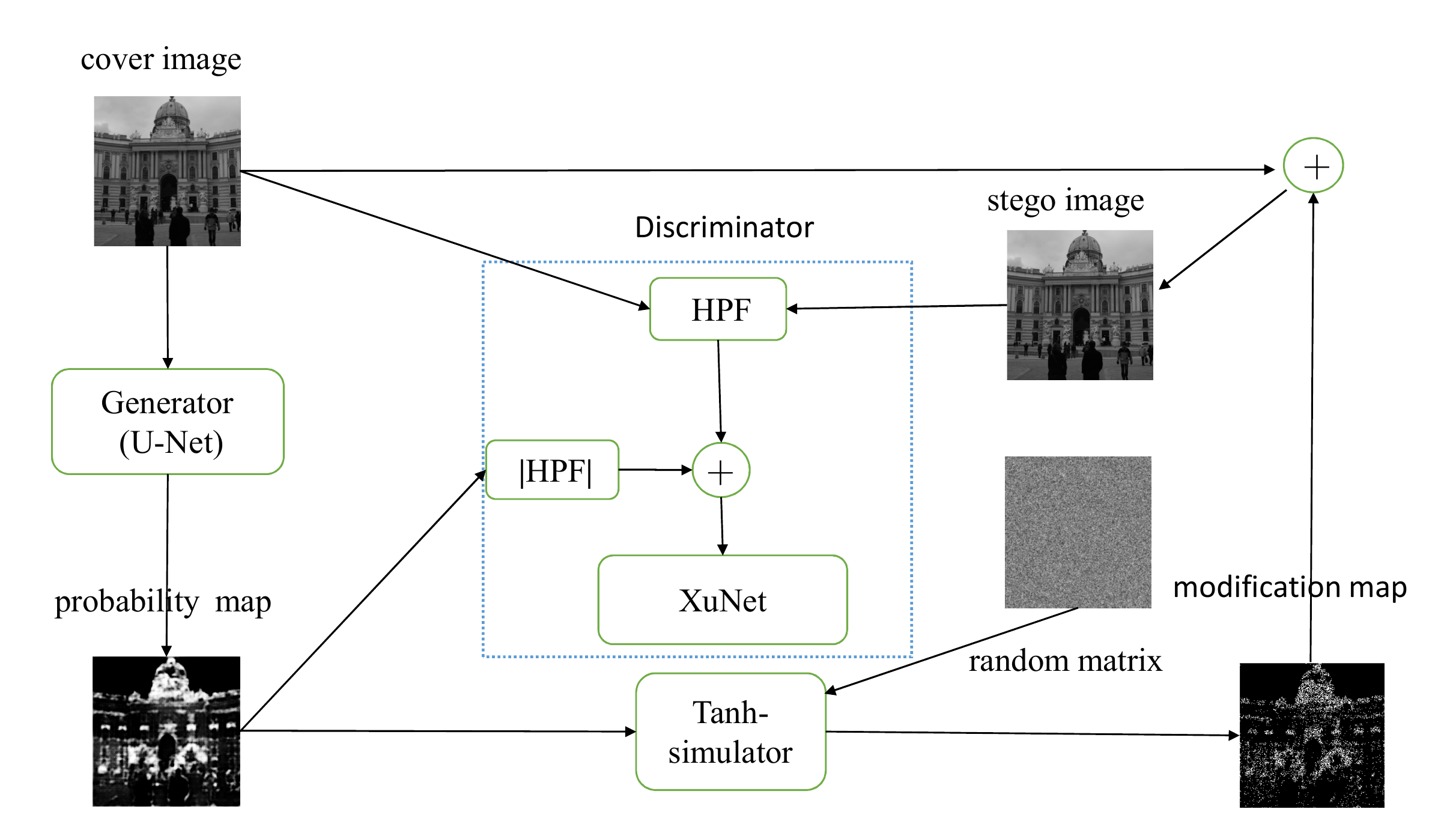}
	\caption{Steganographic architecture of the proposed UT-SCA-GAN. }\label{Architecture}
\end{figure*}

The proposed overall architecture is shown in Fig. \ref{Architecture}. The training steps are described as follows:
\begin{enumerate}[label= (\arabic*)]
	\item  Translate a cover image into an embedding change probability map by using the generator.
	\item  Given an embedding change probability map and a randomly generated matrix with uniform distribution of [0,1], compute the modification map by using the proposed Tanh-simulator.
	\item  Generate the stego image by adding the cover image and its corresponding modification map.
	\item  Feed cover/stego pairs and the corresponding embedding change probability map into the  discriminator.
	\item  Alternately update the parameters of generator and discriminator.
\end{enumerate}
\subsection{The Loss Functions}
The loss function of the discriminator is defined as follows:

\begin{equation}
{l_{D}= - \sum_{i=1}^{2}y_{i}^{'} log(y_{i})}
\end{equation}
where $y_{i}$ is the Softmax output of the discriminator, while $y_{i}^{'}$ is the corresponding truth label of cover/stego.

The loss function of the generator is defined as follows \cite{tang2017automatic}:

\begin{equation}
{l_{G} = -\alpha \times l_{D} + \beta  \times (C - H\times W \times  Q )^{2}}
\end{equation}
where H and W are the height and width of the cover image, Q denotes the embedding payload, and ${C}$ is the capacity that guarantees the payloads:
\begin{equation}
{C = \sum_{i=1}^{H}\sum_{j=1}^{W}-p_{i,j}^{+1}log_{2}p_{i,j}^{+1} -p_{i,j}^{-1}log_{2}p_{i,j}^{-1} - p_{i,j}^{0}log_{2}p_{i,j}^{0}}
\end{equation}
\begin{equation}
{p_{i,j}^{-1} = p_{i,j}^{+1} = p_{i,j}/2}
\end{equation}
\begin{equation}
{p_{i,j}^{0}+p_{i,j}^{-1}+ p_{i,j}^{+1}= 1}
\end{equation}
where $p_{i,j}$ denotes the output embedding probability of the generator corresponding to the pixel $x_{i,j}$, $p_{i,j}^{+1}$ and $p_{i,j}^{-1}$ denote the modify probability of adding or subtracting 1, while $p_{i,j}^{0}$ denote the probability of the corresponding pixel $x_{i,j}$ will not be modified.

\subsection {Generator Design}
\begin{figure}[!htb]
	\centering
	\includegraphics[totalheight=2.7in, width=3.5in, origin=c]{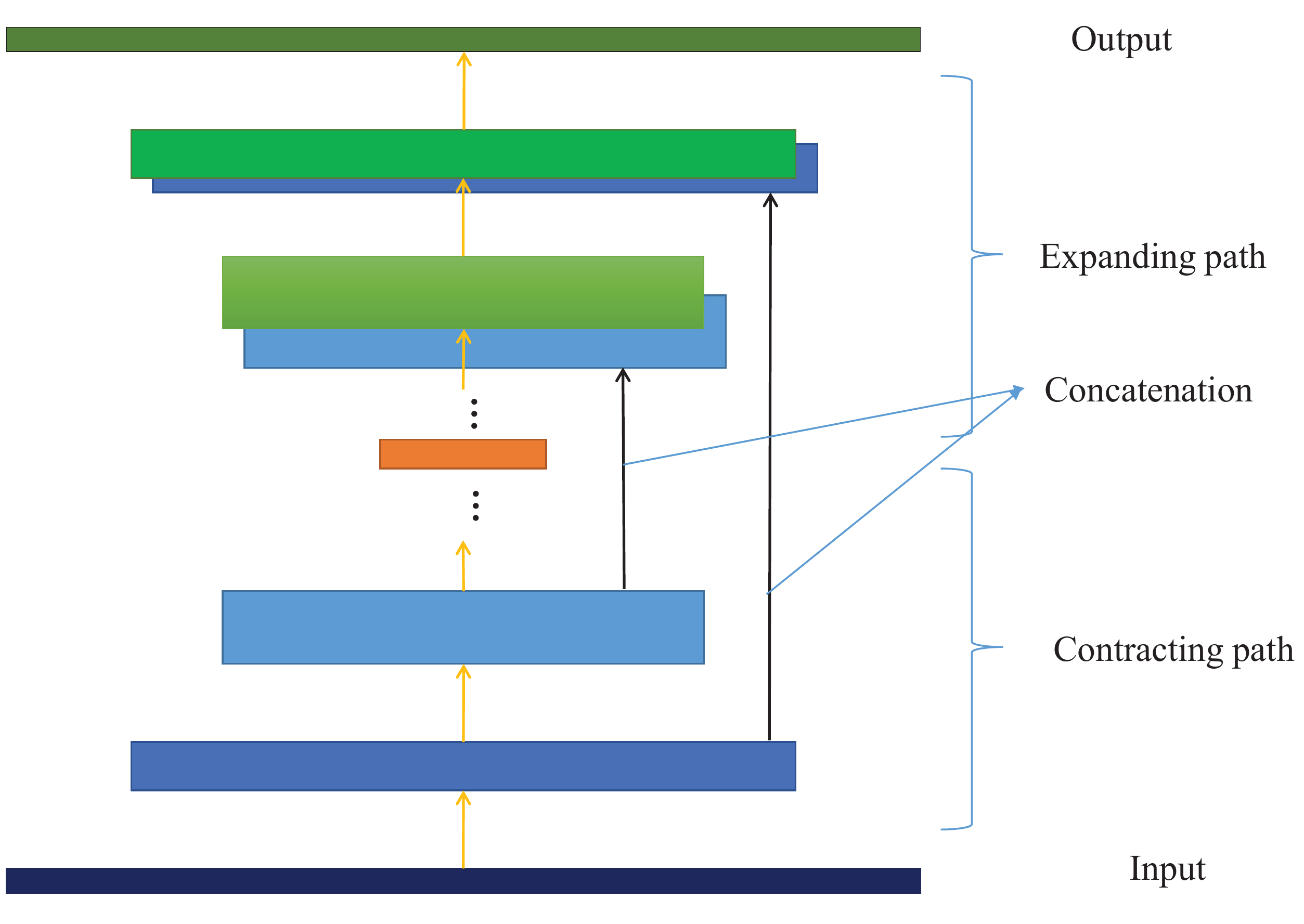}
	\caption{Typical architecture of the U-Net. }\label{U-Net}
\end{figure}

Motivated by an elegant architecture ``U-Net" \cite{UNET}, which was used for biomedical image segmentation, we design an efficient generator for secure steganography based on U-Net. A typical architecture of U-Net is shown in Fig. \ref{U-Net}. The detailed configuration of the proposed generator is shown in Table \ref{tab:generator configuration}. Note that in the contracting path, each group shown in the table corresponds to the sequence of convolution, batch normalization and Leaky-ReLU. A group of the expanding path corresponds to the sequence of deconvolution, batch normalization and ReLU. The last layer ensures that the embedding probability ranges from 0 to 0.5 by considering large embedding probability may caused the embedding process be easily detected \cite{denemark2014further}. The Leaky-ReLU activation function is defined as follows:
\begin{equation}\label{eq2}
f(x)=
\begin{cases}
x&     x>0\\
\alpha x&    x\leqslant 0\\

\end{cases}
\end{equation}

To prevent the ``dying ReLU" problem, we set $\alpha$ = 0.2. The main characteristics of the generator are described as follows:
\begin{enumerate}[label= (\arabic*)]
	
	\item It is composed of the contracting path and the expanding path. The former follows the typical architecture of a convolutional neural network while the latter mainly consists of deconvolution operations.
	\item In order to achieve pixel-level learning and facilitate the back-propagation of gradients, there are concatenation connections between every pair of mirrored layers with the same size, such as layers 1 and 15, layers 2 and 14, etc.
	\item The middle layers capture the global information of the image, while both sides of the generator provide local information.
\end{enumerate}

Different from the 25-layer generator in ASDL-GAN \cite{tang2017automatic}, here the pre-processing layer is not used. In addition, the generator converges quickly and trains faster due to skipping connections and low memory consumption.

\begin {table*}[!htbp]
\caption {Configuration details of the proposed generator.
}\label{tab:generator configuration}
\begin{center}
	\begin{tabular}{ |p{1.5cm}<{\centering}|p{2.5cm}<{\centering}|p{2.5cm}<{\centering}|p{5.0cm}<{\centering}| }
		\hline
		Layers & Output size  & Kernels size & Process                 \\
		\hline
		Input & $1\times(512 \times 512)$ & /      & Convolution-BN-Leaky ReLU   \\
		\hline
		Layer 1 & $16\times(256 \times 256)$ & $16\times(3\times3\times1)$      & Convolution-BN-Leaky ReLU   \\
		\hline
		Layer 2 & $32\times(128 \times 128)$  & $32\times(3\times3\times 16)$       & Convolution-BN-Leaky ReLU   \\
		\hline
		Layer 3 & $64\times(64 \times 64)$  & $64\times(3\times3\times32)$       & Convolution-BN-Leaky ReLU  \\
		\hline
		Layer 4 &$ 128\times(32 \times 32) $ &  $128\times(3\times3\times64)$     &Convolution-BN-Leaky ReLU    \\
		\hline
		Layer 5 &$ 128\times(16 \times 16)$  & $128\times(3\times3\times128)$        & Convolution-BN-Leaky ReLU  \\
		\hline
		Layer 6 & $128\times(8 \times 8) $   & $128\times(3\times3\times128)$      & Convolution-BN-Leaky ReLU   \\
		\hline
		Layer 7 & $128\times(4 \times 4)$  &   $128\times(3\times3\times128)$     & Convolution-BN-Leaky ReLU   \\
		\hline
		Layer 8 & $128\times(2 \times 2) $  & $128\times(3\times3\times128)$      & Convolution-BN-Leaky ReLU   \\
		\hline
		Layer 9 & $256\times(4 \times 4)$ &  $128\times(5\times5\times128)$      & Deconvolution-BN-ReLU-Concatenate   \\
		\hline
		Layer 10 & $256\times(8 \times 8)$ &   $128\times(5\times5\times128)$      &Deconvolution-BN-ReLU-Concatenate    \\
		\hline
		Layer 11 & $256\times(16 \times 16) $  &  $128\times(5\times5\times256)$     &Deconvolution-BN-ReLU-Concatenate    \\
		\hline
		Layer 12 & $256\times(32 \times 32)$  &  $128\times(5\times5\times256)$    & Deconvolution-BN-ReLU-Concatenate    \\
		\hline
		Layer 13 & $128\times(64 \times 64)$ &   $64\times(5\times5\times256)$     & Deconvolution-BN-ReLU-Concatenate    \\
		\hline
		Layer 14 & $64\times(128 \times 128) $ & $32\times(5\times5\times128)$      & Deconvolution-BN-ReLU-Concatenate     \\
		\hline
		Layer 15 & $32\times(256 \times 256) $ & $16\times(5\times5\times64)$    & Deconvolution-BN-ReLU-Concatenate    \\
		\hline
		Layer 16 & $1\times(512 \times 512)$  & $1\times(5\times5\times32)$     & Deconvolution-BN-ReLU-Concatenate    \\
		\hline
		Layer 17 & $1\times(512 \times 512)$  &   /  & ReLU((Sigmoid-0.5))   \\
		\hline
		
	\end{tabular}
\end{center}
\end{table*}

\subsection {Tanh-simulator Function}
In the previous adaptive steganography methods  \cite{HUGO,WOW,HILL,UNIWARD,MiPOD}, stego image is generated by adding the cover image and the corresponding modification map. The modification map is computed by using an optimal embedding simulator \cite{Simulator}, which is a staircase function:

\begin{equation}\label{eq1}
{m}_{i,j}=\begin{cases}
-1& \text{if } {n}_{i,j}< {p}_{i,j}/2\\
1& \text{if } {n}_{i,j}>1- {p}_{i,j}/2 \\
0&  \text{otherwise.}
\end{cases}
\end{equation}
where ${p}_{i,j}$ is the embedding change probability, ${n}_{i,j}$ is the random number generated from a uniform distribution between 0 and 1, ${m}_{i,j}$ is the embedding value.

\begin{figure}
	\begin{subfigure}[t]{0.25\textwidth}
		\centering
		\includegraphics[width=\linewidth]{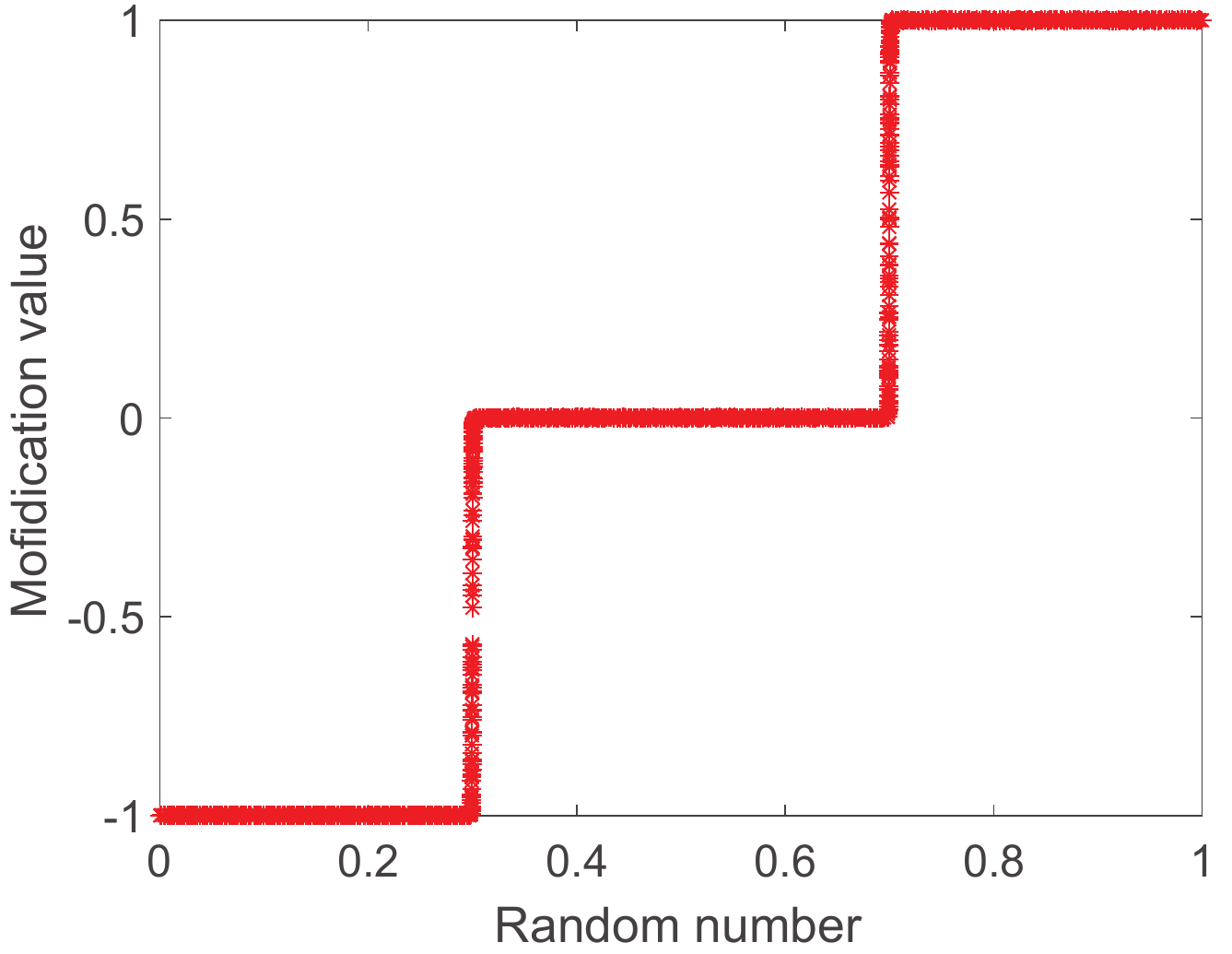}
		\caption{Tanh-simulator ($\lambda $=1000)}
		\label{fig:side:a}
	\end{subfigure}\hspace*{\fill}
	\begin{subfigure}[t]{0.25\textwidth}
		\centering
		\includegraphics[width=\linewidth]{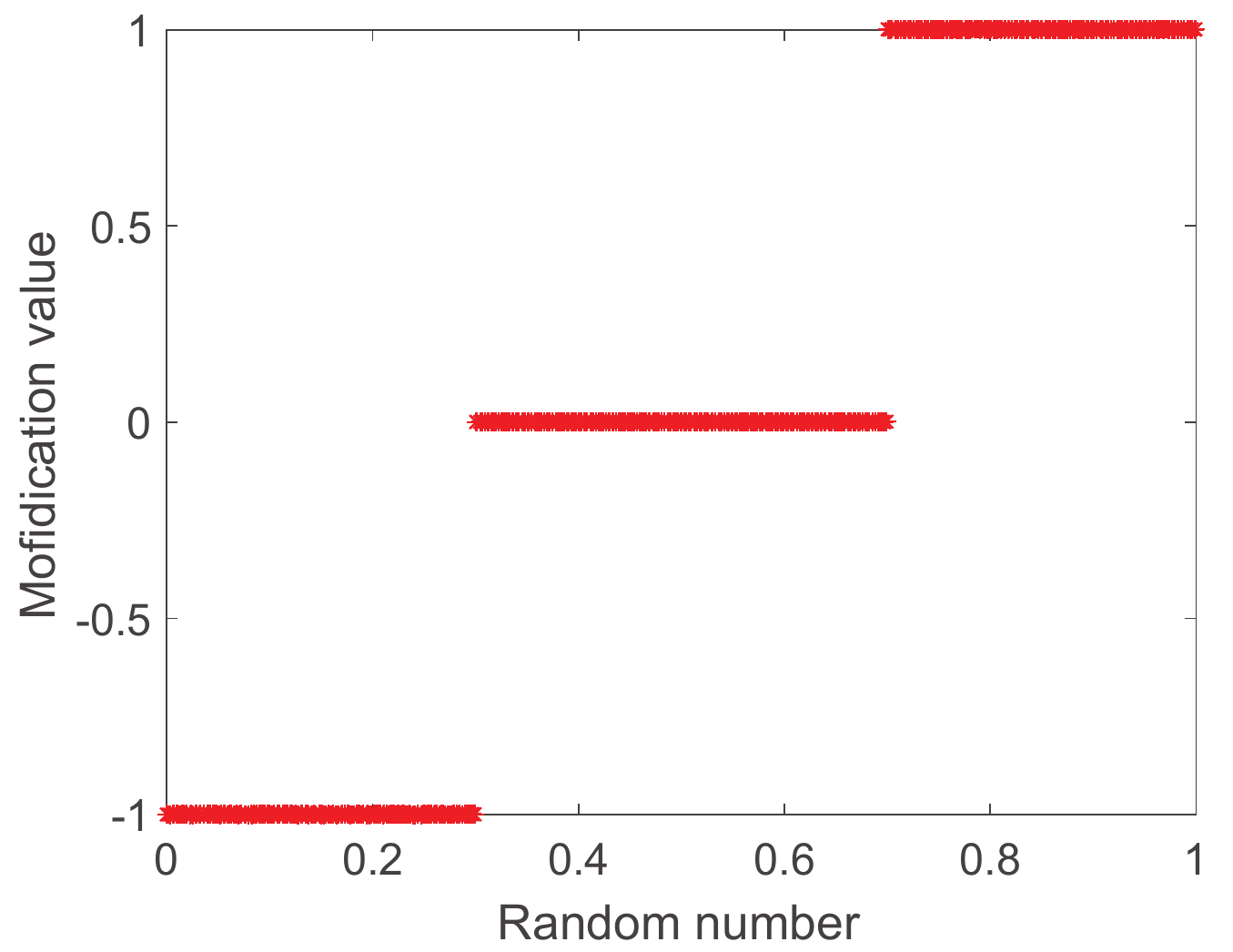}
		\caption{embedding simulator \cite{Simulator}}
		\label{fig:side:b}
	\end{subfigure}
	
	\caption{Function curves of the Tanh-simulator and the embedding simulator in 2D space (${p}_{i,j}$= 0.6): (a) Tanh-simulator ($\lambda $=1000), (b) embedding simulator \cite{Simulator}.}\label{fig:tanh-2d-simulator}
\end{figure}

\begin{figure}
	\begin{subfigure}[t]{0.25\textwidth}
		
		\centering
		\includegraphics[width=\linewidth]{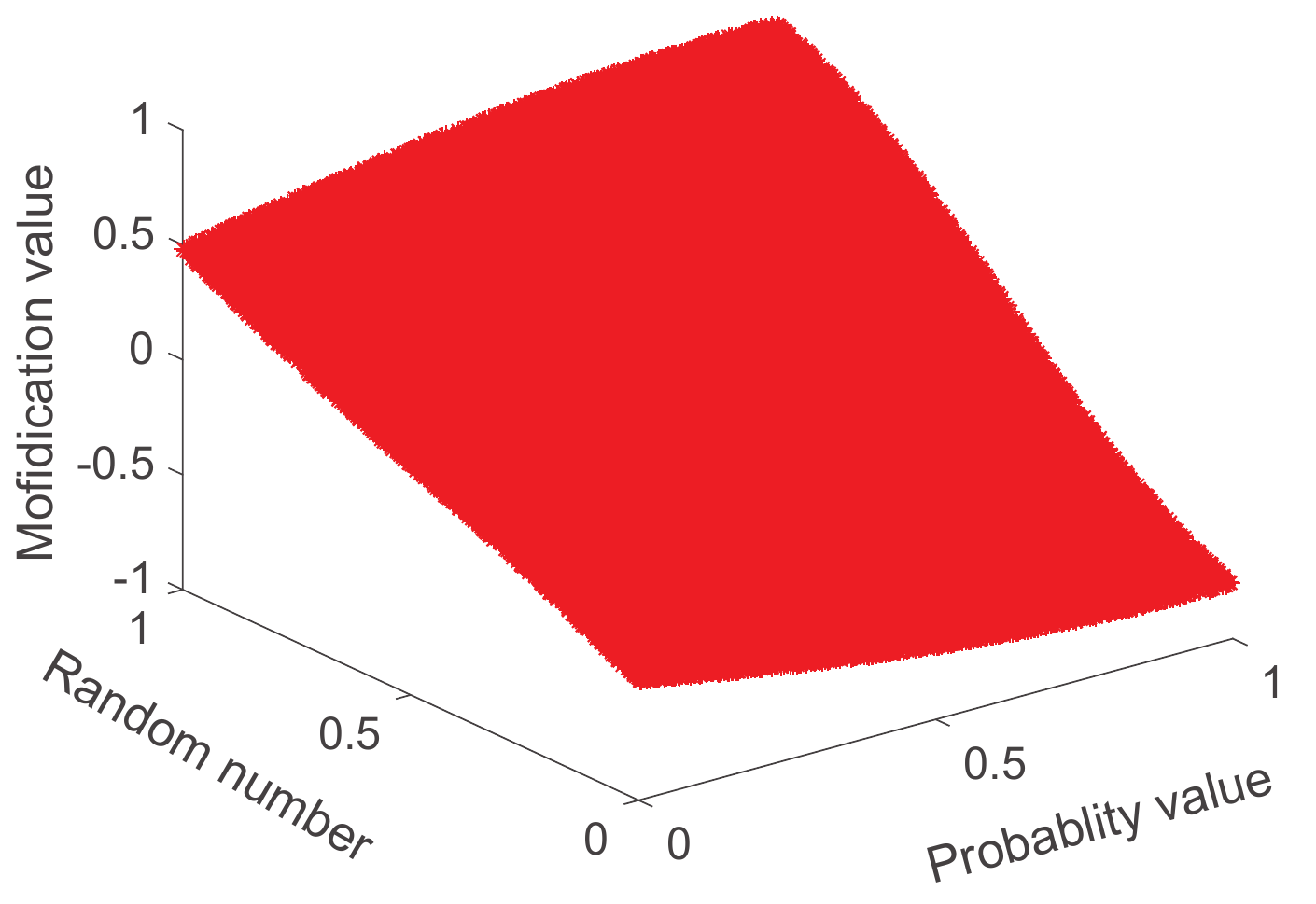}
		\caption{}
		\label{fig:side:a}
	\end{subfigure}\hspace*{\fill}
	\begin{subfigure}[t]{0.25\textwidth}
		\centering
		\includegraphics[width=\linewidth]{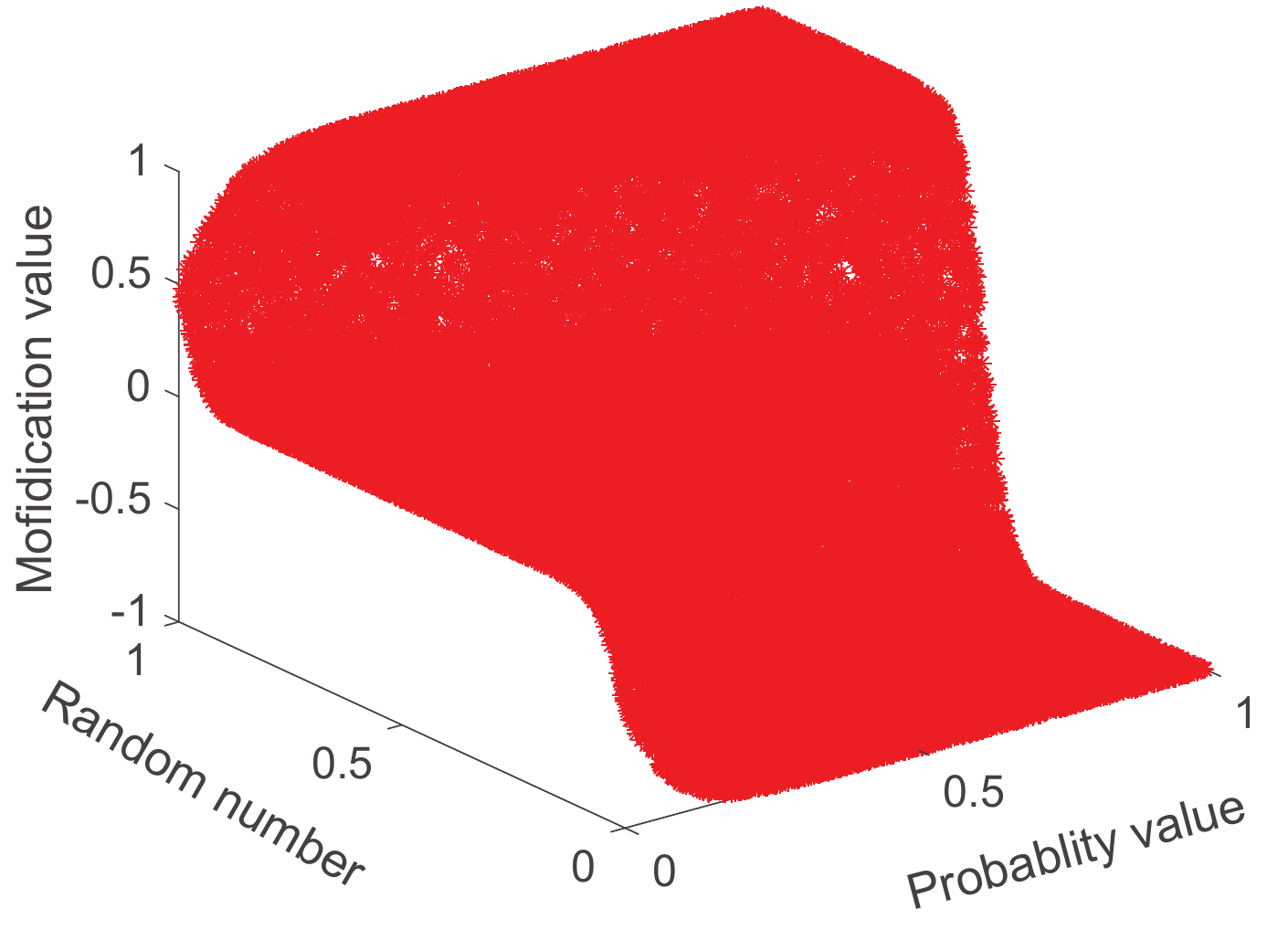}
		\caption{}
		\label{fig:side:b}
	\end{subfigure}
	
	\begin{subfigure}[t]{0.25\textwidth}
		\centering
		\includegraphics[width=\linewidth]{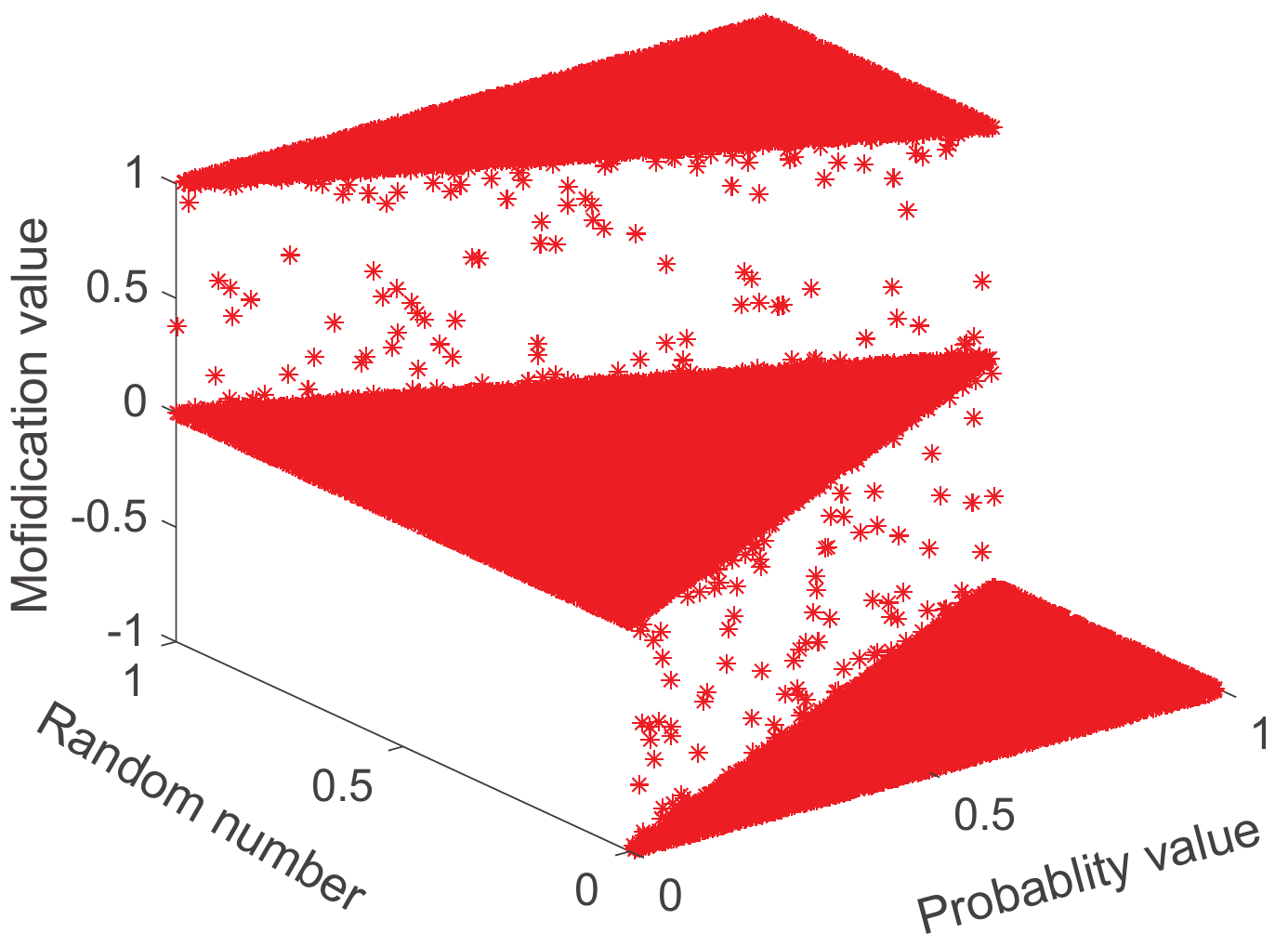}
		\caption{}
		\label{fig:side:a}
	\end{subfigure}\hspace*{\fill}
	\begin{subfigure}[t]{0.25\textwidth}
		\centering
		\includegraphics[width=\linewidth]{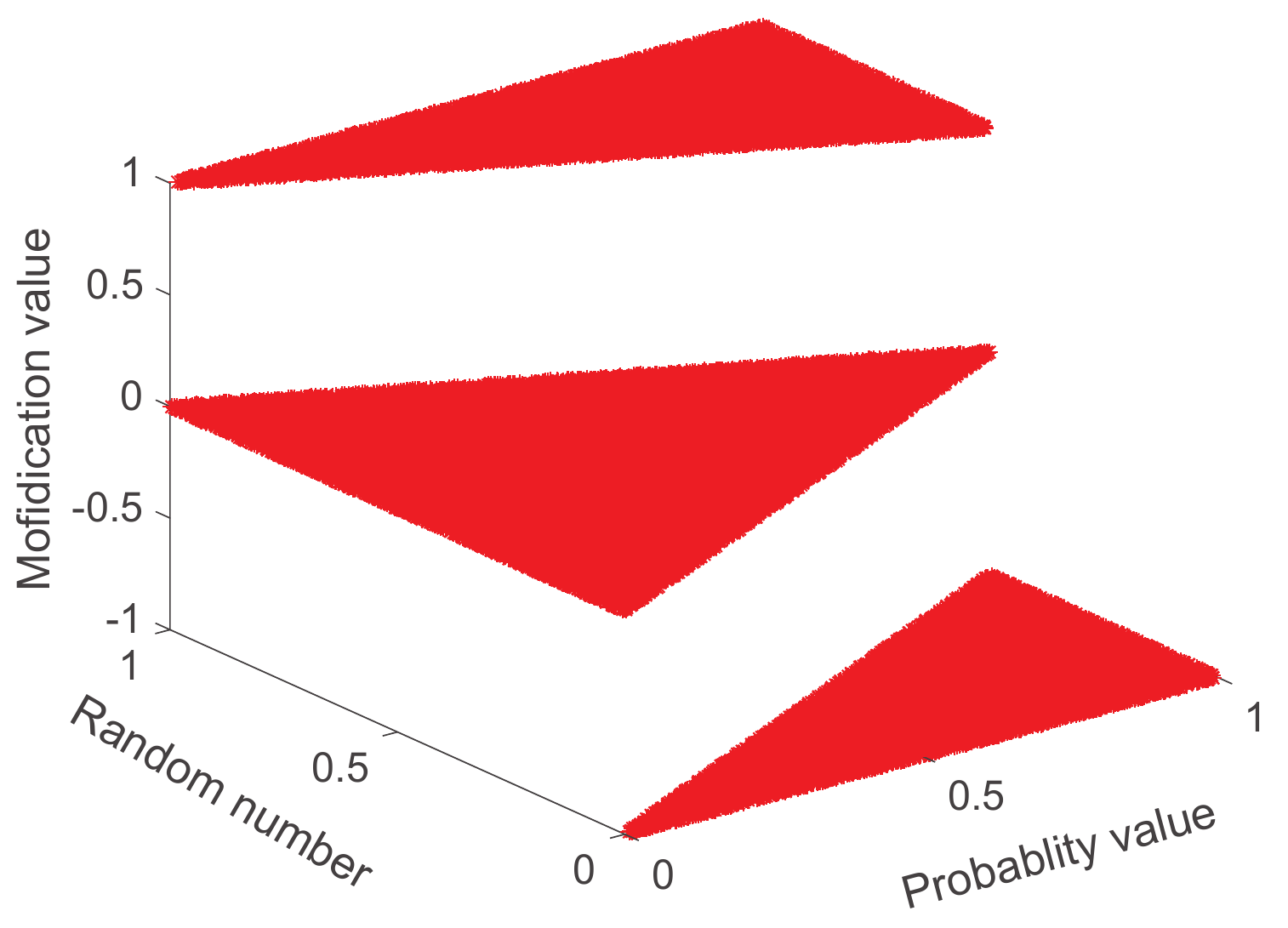}
		\caption{}
		\label{fig:side:b}
	\end{subfigure}
	\caption{Comparison between Tanh-simulator and embedding simulator in 3D space: (a) Tanh-simulator ($\lambda$=1), (b) Tanh-simulator ($\lambda$=10), (c) Tanh-simulator ($\lambda$=1000), (d) embedding simulator \cite{Simulator}. }\label{fig:tanh}
\end{figure}

Because the staircase function (\ref{eq1}) cannot convey the gradient during back propagation, we use the Tanh function to fit the embedding simulator. We called the proposed activation function Tanh-simulator. It can be described as follows:



\begin{equation}\label{eq2}
\begin{split}
{m}'_{i,j}=-0.5 \times tanh\left ( \lambda  \left ( {p}_{i,j}-2\times {n}_{i,j} \right ) \right )+ \ \\
0.5 \times tanh\left ( \lambda \left ({p}_{i,j}-2 \times \left ( 1- {n}_{i,j}\right ) \right )\right )
\end{split}
\end{equation}

\begin{equation}\label{eqtanh}
tanh \left ( x \right ) =   \frac{e^{x}-e^{-x}}{e^{x}+e^{-x}}
\end{equation}
where $\lambda $ is the scaling factor, which controls the slope at the junction of stairs. Fig. \ref{fig:tanh-2d-simulator} and  Fig. \ref{fig:tanh} illustrate the function curves of the Tanh-simulator and the staircase function (\ref{eq1}) in 2D and 3D respectively. It can be seen that with the increment of parameter $\lambda $, Tanh-simulator is becoming more similar to the staircase function (\ref{eq1}). Note that we need some discrete points to convey the gradient, considering the conveyance of gradient and fitting accuracy, wet set $\lambda $ = 1000.

\subsection {Discriminator Design}

Discriminator is a steganalytic tool in this framework. Considering the adversarial training
between steganography and steganalysis, we infer that the enhancement of the discriminator will certainly
force the steganography to be more secure.

To resist the current SCA based steganalysis method, we incorporate SCA into the discriminator.
The $\left|HPF\right|$ in Fig. \ref{Architecture} denote the absolute value of the 30 high pass
filters from SRM to consider the statistical measure \cite{Ni2017}. Through adversarial training,
the generator will adjust the parameters to resist the SCA based discriminator. In addition,
the embedding change probability map bypasses the Tanh-simulator will accelerate the gradient propagation in adversarial training.

\section{Experimental results and analysis}

\subsection {Experimental Setting}
All experiments are conducted on the SZUBase \cite{tang2017automatic} and BOSSBase-1.01 \cite{BOSSBase} which contains images with size of 512 $\times$ 512. The first dataset that contains 40,000 grayscale cover images is used to train the proposed architecture. The second dataset that contains 10,000 grayscale images is used to generate stego images. 5,000 pairs of images from BOSSBase are randomly selected to train the ensemble classifier \cite{Ensemble}, and the left 5,000 pairs are used as the test set. We use the Adam optimizer with the learning rate of 0.0001 to train the model over 160,000 iterations (32 epochs). During training, 8 cover/stego pairs are used as input in each iteration. The parameters  $\alpha$, $\beta$ are set to 1, and 10$^{-7}$, respectively. All experiments are performed with TensorFlow on a GTX 1080 Ti GPU card.

\subsection {Experiments on resized dataset}
In this part, we conduct experiments to investigate the influence of different parts of the proposed architecture. The original dataset of SZUBase and BOSSbase-1.01 are resized to 256 $\times$ 256 by ``imresize()" function in Matlab.

Firstly, we conduct experiments on UT-GAN, which is the variant of UT-SCA-GAN without SCA. We vary the UT-GAN as follows:
\begin{enumerate}[label= (\arabic*)]
	\item Variant \#1, replace the proposed generator with the generator of ASDL-GAN.
	\item Variant \#2, replace the Tanh-Simulator with the TES network of ASDL-GAN.
\end{enumerate}

Experimental results tested on BOSSBase 256 $\times$ 256 are shown on Table \ref{tab:Var_UT_GAN}. All of the methods are embedded messages with payload 0.4 bpp. From Table \ref{tab:Var_UT_GAN}, it can be seen that the performance will decrease dramatically if we replace the generator or embedding simulator with the corresponding parts of the ASDL-GAN. Replacement of the generator caused the greatest performance reduction, it is because the proposed generator directly determines the adaptability and security of the steganographic scheme. The Tanh-simulator also achieves better performance than the TES network, the result might be caused by high fitting accuracy. In addition, the proposed UT-GAN also obtains better performance than ASDL-GAN and S-UNIWARD.

\begin {table*}[!htb]
\caption {Error rates (\%) of different steganographic schemes detected by SRM \cite{SRM} on BOSSbase 256 $\times$ 256.
}\label{tab:Var_UT_GAN}
\begin{center}
	\begin{tabular}{|p{1.5cm}<{\centering}|p{1.5cm}<{\centering}|p{1.5cm}<{\centering}|p{1.5cm}<{\centering}|p{2.0cm}<{\centering}|p{2.0cm}<{\centering}|}
		\hline
		Algorithm & UT-GAN             &Variant \#1   & Variant \#2 & ASDL-GAN \cite{tang2017automatic}&  S-UNIWARD\cite{UNIWARD}   \\
		\hline
		Error rates& \textbf{26.61}      &  20.29            &  23.06 & 20.68           &   22.26             \\
		\hline
	\end{tabular}
\end{center}
\end {table*}

Next, we conduct experiments with the UT-SCA-GAN which incorporates SCA in the discriminator to verify the influence of SCA. The payload of  UT-SCA-GAN and UT-GAN are set as 0.4 bpp. We test the performance by SRM and maxSRMd2. Experimental results are show on Table \ref{tab:test_sca}.

As we expected, the incorporation of SCA to the discriminator will improve the security performance by about 2.0\% to resist SCA based steganalysis method, e.g. maxSRMd2. Because the SCA is incorporated into the discriminator, the parameters of generator can be automatically adjusted according to the structure of discriminator via the adversarial training.

\begin {table}[!htb]
\caption {Error rates (\%) of UT-GAN and UT-SCA-GAN on BOSSBase 256 $\times$ 256.
}\label{tab:test_sca}
\begin{center}
\begin{tabular}{|p{2.4cm}<{\centering}|p{2.4cm}<{\centering} |p{2.4cm}<{\centering}|}
		\hline
		Net-work& maxSRMd2 &SRM   \\
		\hline
		UT-GAN  & 20.27 & 26.61       \\
		\hline
		UT-SCA-GAN  &  \textbf{22.30} &  26.43       \\
		\hline
\end{tabular}
\end{center}
\end {table}

\subsection {Experiments on full size dataset}
In this part, we conduct experiments on size of 512 $\times $512 to compare with previous method. For 0.1 bpp, we finetune the architecture from the trained model with 0.4 bpp. We find this process will improves the security performance by about 1.0\%. For 0.2 bpp, we only compare with S-UNIWARD because work \cite{tang2017automatic} did not run experiment on this payload. It is observed from Table \ref{tab:Compare with tang 512} that the proposed UT-GAN performs better than  ASDL-GAN by 4.96\% and 7.80\% for 0.4 bpp and 0.1 bpp respectively. It also obtained better performance than S-UNIWARD. From Table \ref{tab:compare_SCA}, the incorporation of SCA can also improve the performance on full size image of 512 $\times $512. The performance improvement of incorporation of SCA will be more significantly with the increment of payload. This is may because of the hard training for low payload by using deep-leaning based method, no matter steganalysis or steganography.

\begin {table}[!htb]
\caption {Error rates (\%) of different steganographic schemes detected by SRM \cite{SRM} on BOSSBase 512 $\times$ 512.
}\label{tab:Compare with tang 512}
\begin{center}
	\begin{tabular}{|p{1.4cm}<{\centering}|p{1.4cm}<{\centering}|p{2.0cm}<{\centering}|p{2.0cm}<{\centering}|}
		\hline
		Payload & UT-GAN  &ASDL-GAN \cite{tang2017automatic}  & S-UNIWARD \cite{UNIWARD}    \\
		\hline
		0.4 bpp & \textbf{22.36}      & 17.40        & 20.54\\
		\hline
		0.2 bpp & \textbf{33.03}      & /        & 31.89\\
		\hline
		0.1 bpp& \textbf{40.82}       & 33.02        & 40.40\\
		\hline
	\end{tabular}
\end{center}
\end {table}
\begin {table}[!htbp]
\caption {Error rates  (\%) of different steganographic schemes detected by maxSRMd2 \cite{maxSRM}.
}\label{tab:compare_SCA}
\begin{center}
	\begin{tabular}{|p{2.4cm}<{\centering}|p{2.4cm}<{\centering} |p{2.4cm}<{\centering}|}
		\hline
		Payload &UT-SCA-GAN& UT-GAN       \\
		\hline
		0.4 bpp &\textbf{20.42}&18.23  \\
		\hline
		0.2 bpp &\textbf{28.04}& 26.87  \\
		\hline
		0.1 bpp &\textbf{34.89}&34.64  \\
		\hline
	\end{tabular}
\end{center}
\end {table}

In addition, we also compare the training time of UT-GAN and ASDL-GAN in one epoch. ASDL-GAN needs 4.65 hours, while UT-GAN only needs 1.30 hours. Thus the proposed method saves almost 100 hours for 32 epoch (41.6 hours vs 148.8 hours). There are two reasons for this difference. One reason is the simple architecture of proposed generator, as opposed to the 25-layer generator of the ASDL-GAN. What's more, the time consumption of the proposed Tanh-Simulator function is negligible compared to the two independent 4-layer TES sub-network of ASDL-GAN.

\begin{figure*}[!htbp]
	
	\begin{subfigure}[t]{0.2\textwidth}
		\centering
		\includegraphics[width=\linewidth]{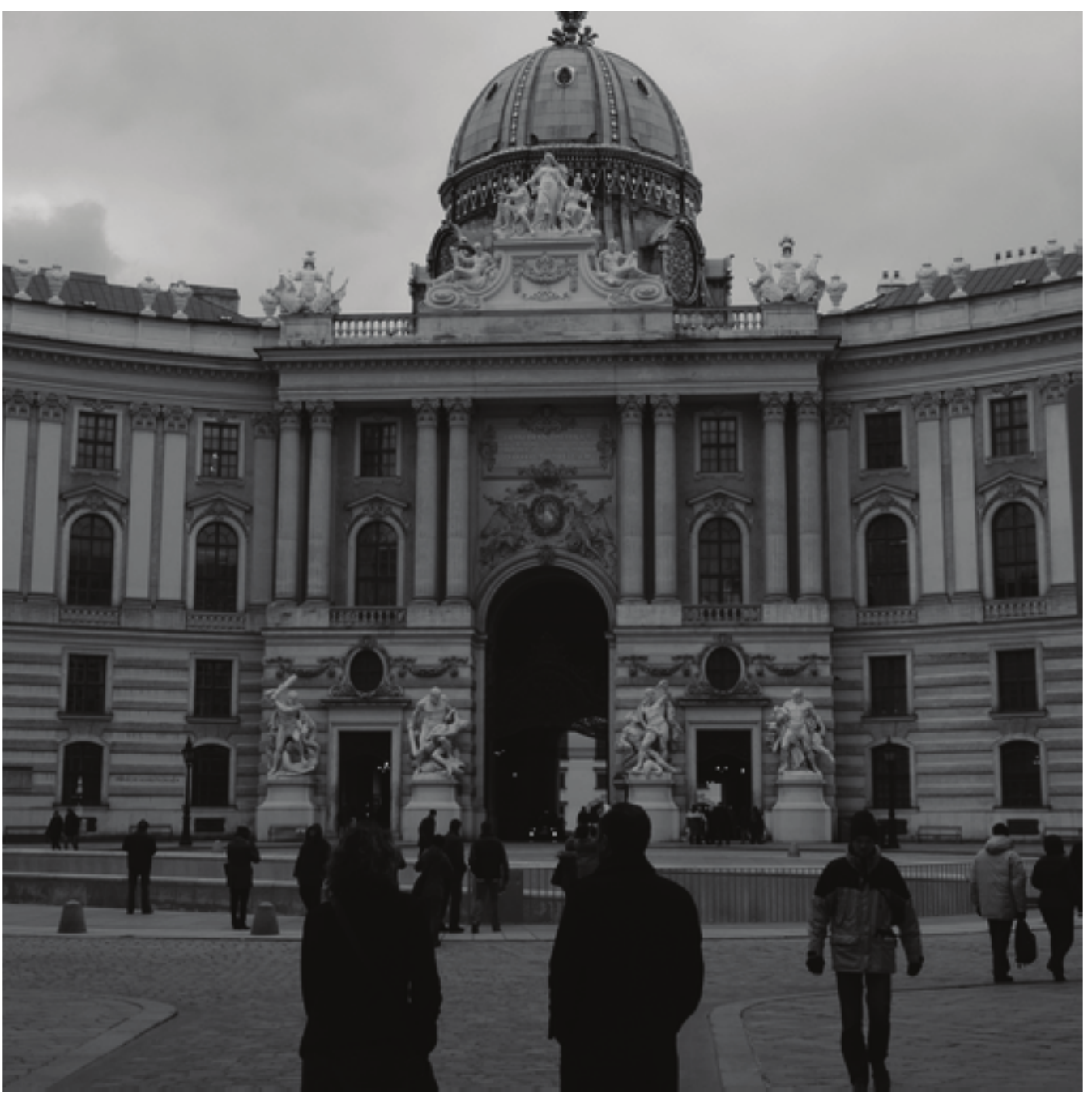}
		\caption{}
		\label{fig:side:a}
	\end{subfigure}\hspace*{\fill}
	\begin{subfigure}[t]{0.2\textwidth}
		\centering
		\includegraphics[width=\linewidth]{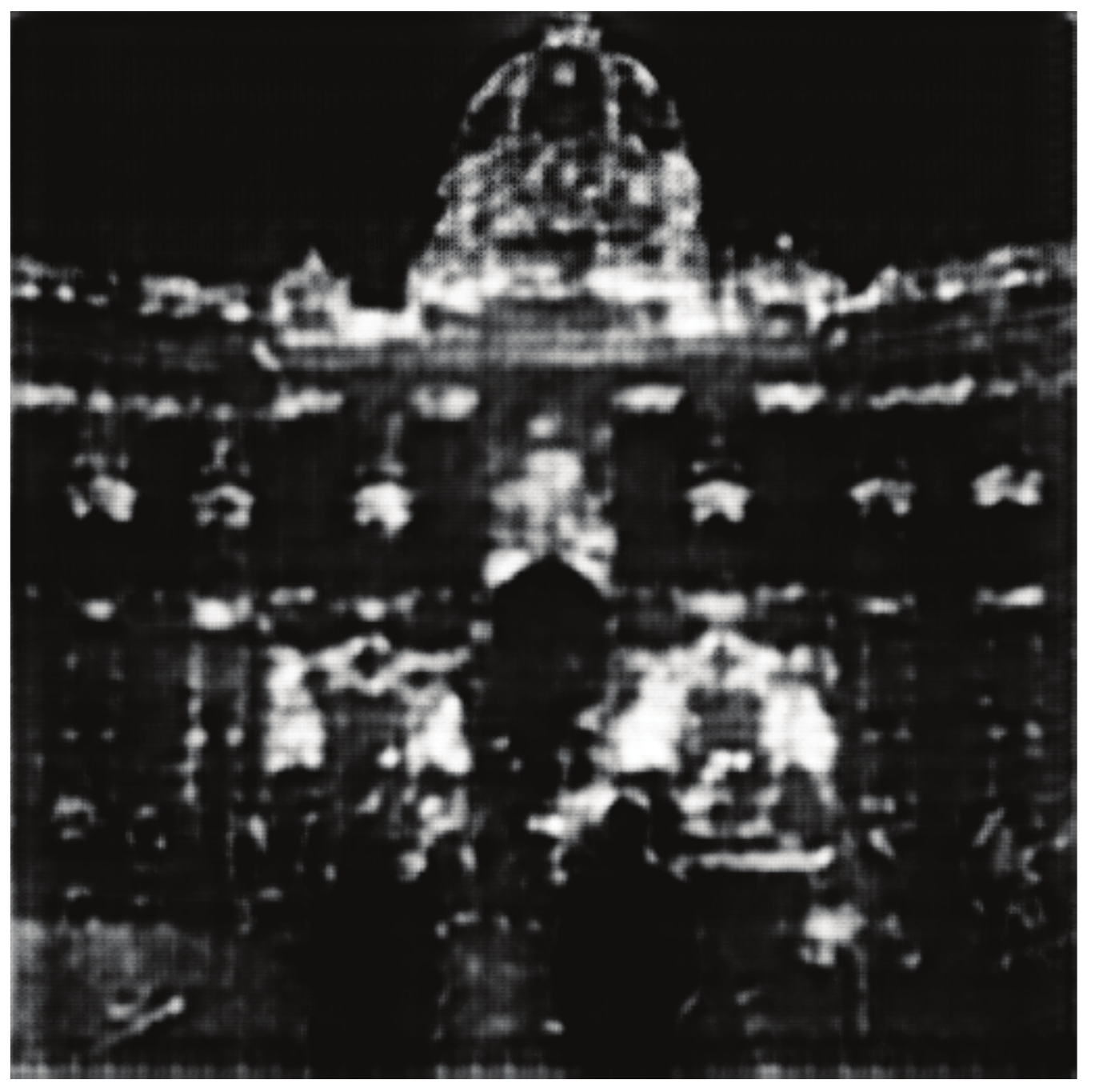}
		\caption{}
		\label{fig:side:a}
	\end{subfigure}\hspace*{\fill}
	\begin{subfigure}[t]{0.2\textwidth}
		\centering
		\includegraphics[width=\linewidth]{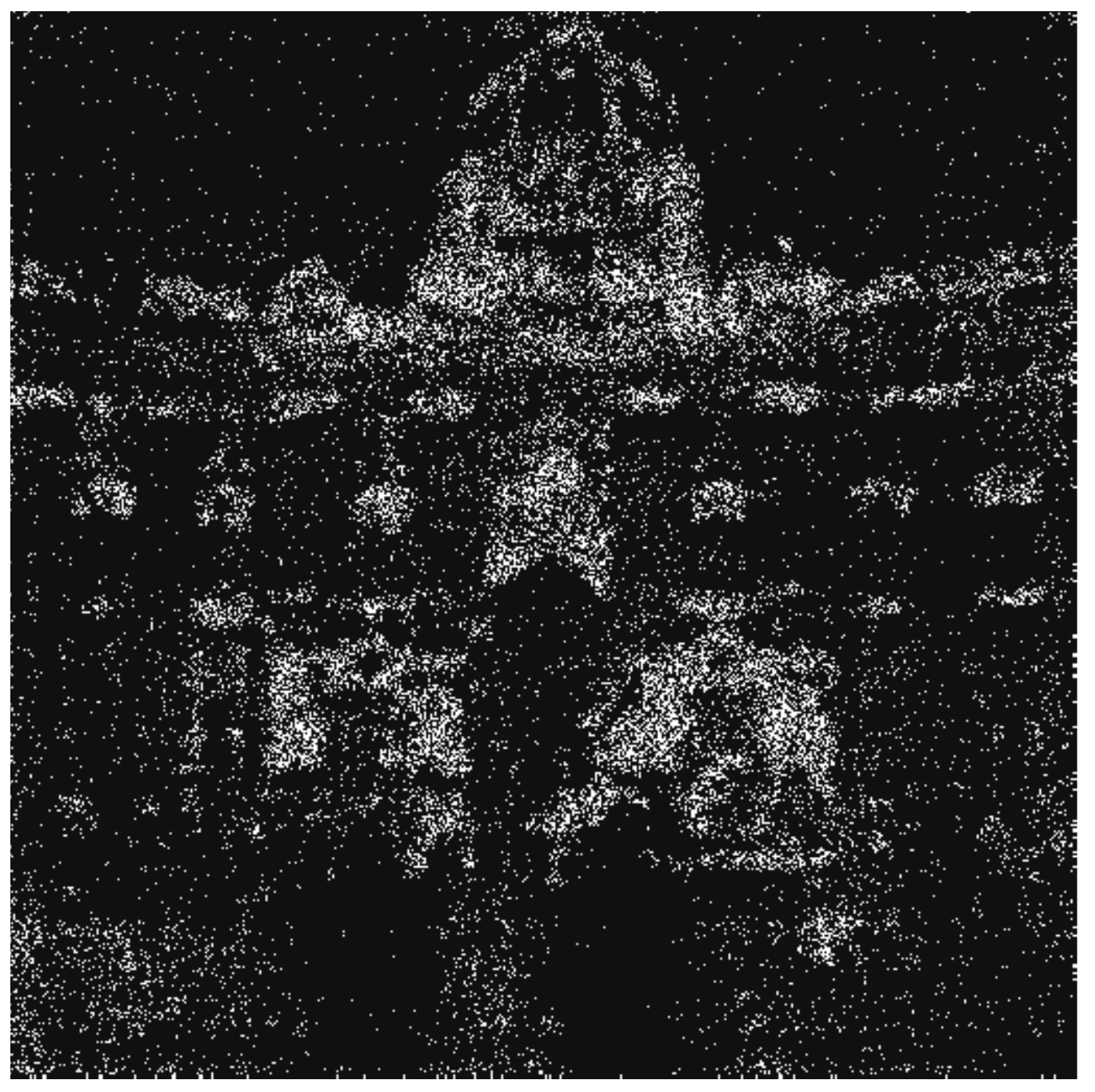}
		\caption{}
		\label{fig:side:a}
	\end{subfigure}\hspace*{\fill}
	\begin{subfigure}[t]{0.2\textwidth}
		\centering
		\includegraphics[width=\linewidth]{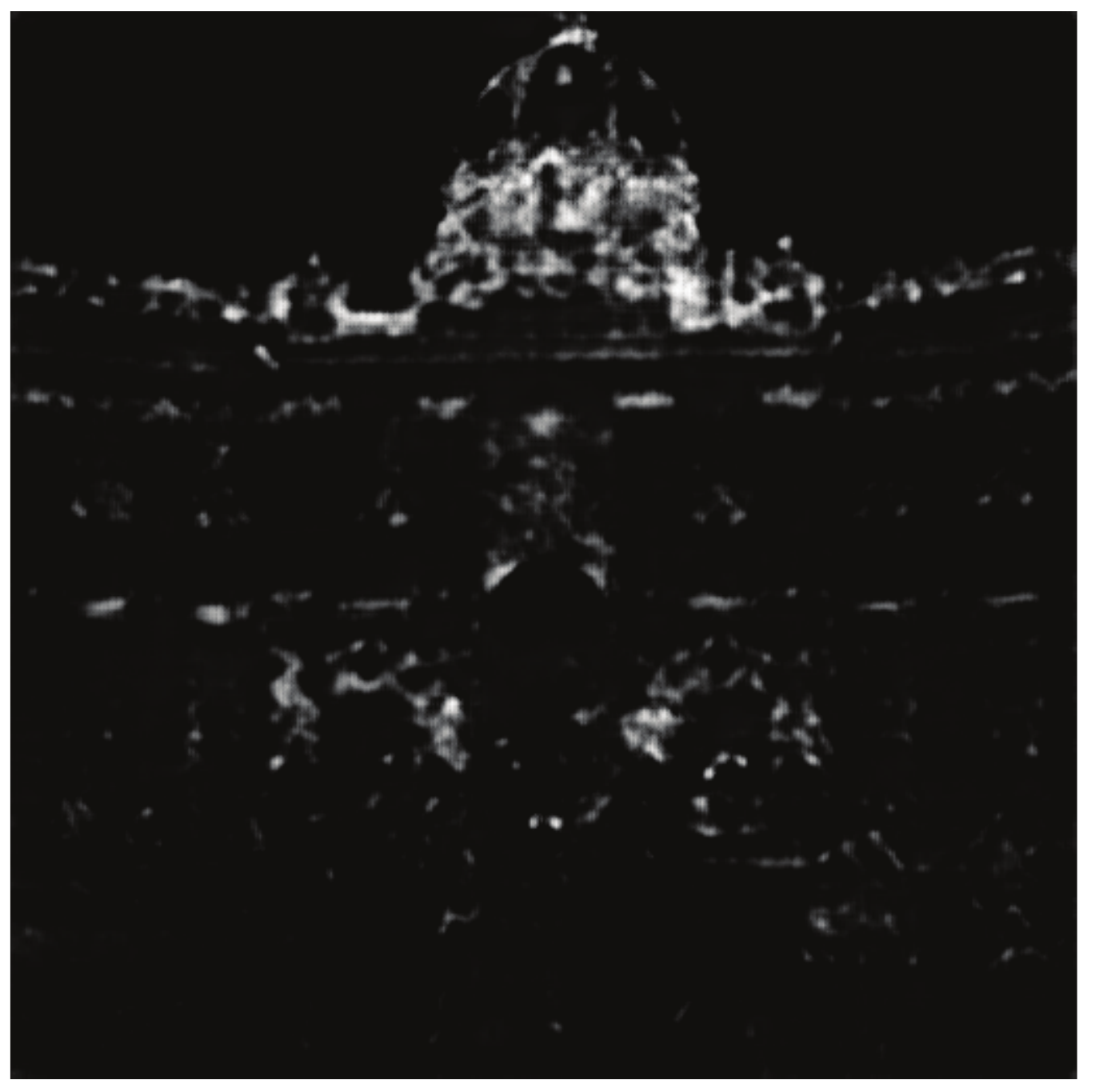}
		\caption{}
		\label{fig:side:a}
	\end{subfigure}\hspace*{\fill}
	\begin{subfigure}[t]{0.2\textwidth}
		\centering
		\includegraphics[width=\linewidth]{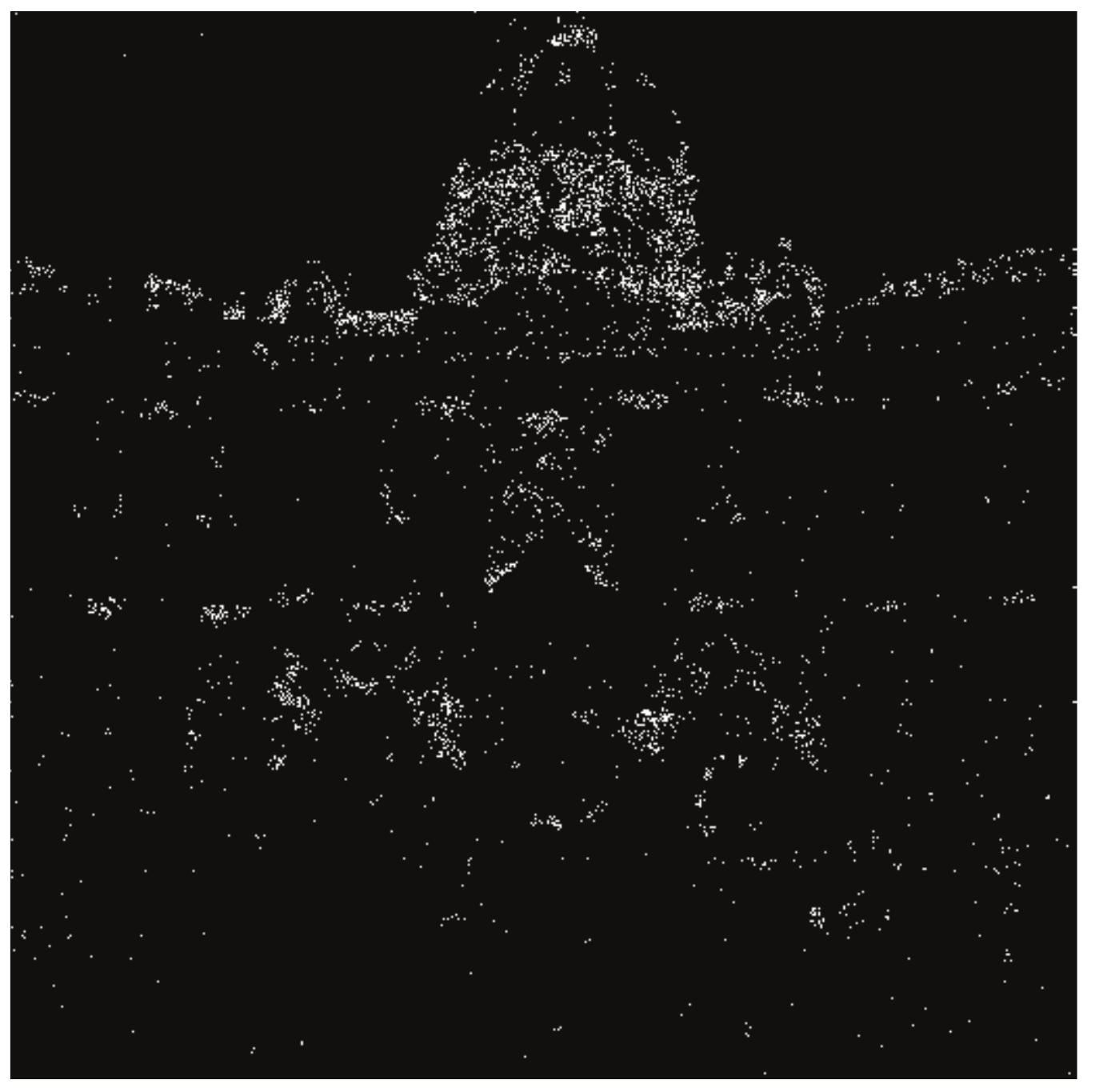}
		\caption{}
		\label{fig:side:a}
	\end{subfigure}\hspace*{\fill}
	
	\caption{Illustration of the proposed UT-GAN. (a) The BOSSBase cover image ``1013.pgm" with a size of  512 $\times$ 512, (b) embedding change probability (0.4 bpp), (c) modification  map (0.4 bpp), (d) embedding change probability map (0.1 bpp), (e) modification  map (0.1 bpp).} \label{fig:compare modification}
	
\end{figure*}

We present the embedding change probability maps and modification position maps with payloads of 0.4 bpp and 0.1 bpp in Fig. \ref{fig:compare modification}. From (b) and (d), it can be seen that embedding change probability value of the texture regions are larger than the smooth regions. From (c) and (e), the embedding change position are also concentrated on regions with large embedding change probability values. Fig. \ref{fig:compare modification} shows that the proposed steganography scheme is content-adaptively.

\section {Practical application of the proposed architecture}
In this work, our task is to learn the embedding probability $p_{i,j}$ by adversarial training. For every pixel of the cover image, the proposed Tanh-simulator has been used to simulate the embedding process. In a practical application scenario, it is necessary to compute the embedding cost by taking full use the learned embedding probability, and then using the Syndrome Trellis codes \cite{STC} to embed the secret information. The embedding cost $\rho_{ij}$ for the practical steganographic coding scheme can be computed as follows \cite {MiPOD}:
\begin{equation}
\rho _{ij} = ln (1/p_{i,j}-2))
\end{equation}

\begin{figure*}[!htb]
	\centering
	\includegraphics[totalheight=2.7in, width=5.5in, origin=c]{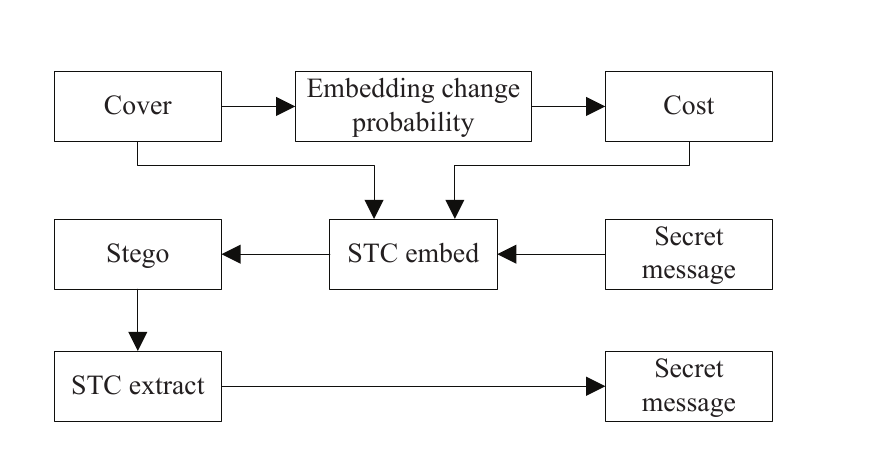}
	\caption{Flowchart of practical embedding process. }\label{fig:STC flowchart}
\end{figure*}

\begin{figure*}[!tbp]
	\begin{subfigure}[t]{0.2\textwidth}
		\centering
		\includegraphics[width=\linewidth]{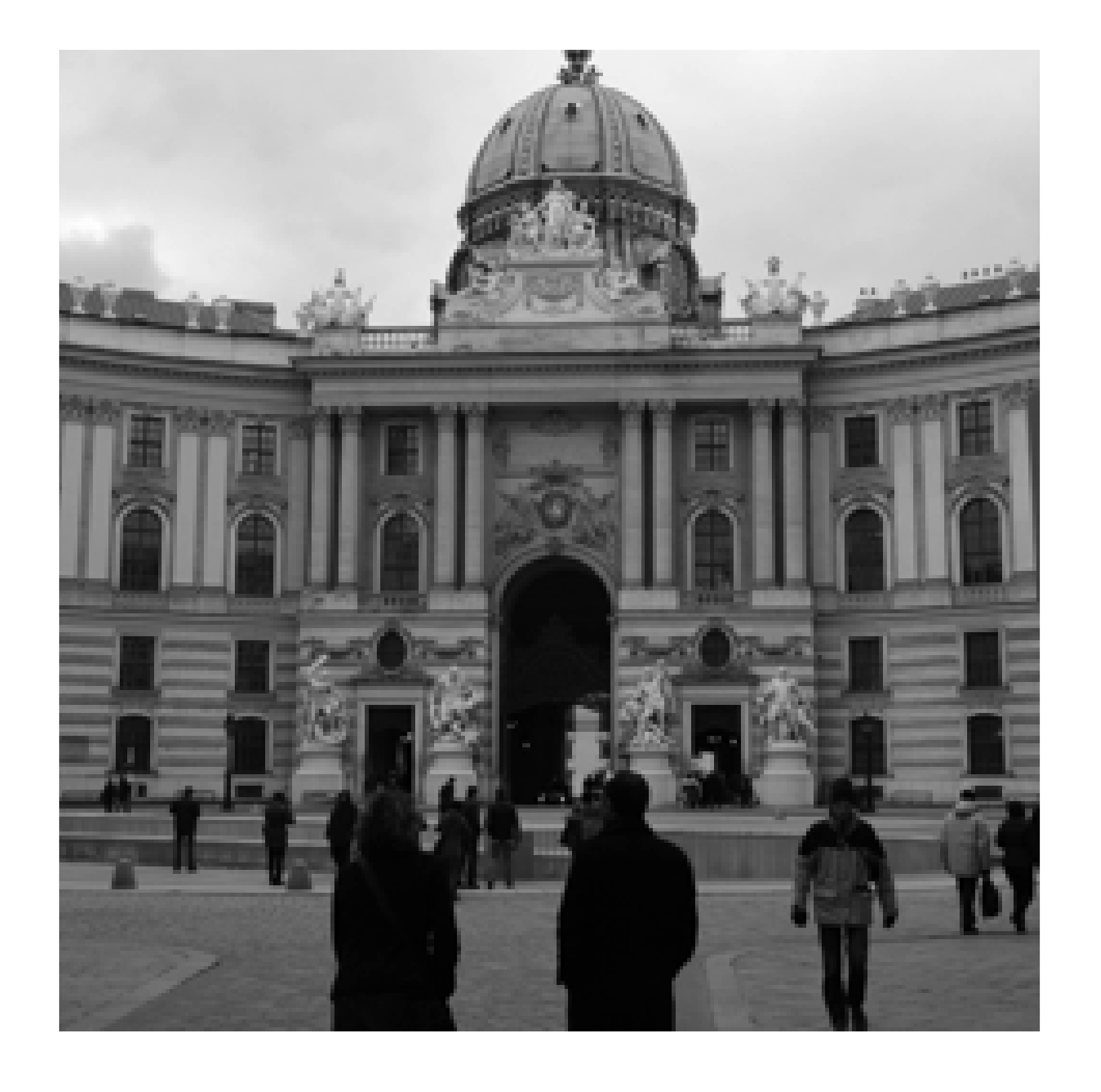}
		\caption{}
		\label{fig:side:a}
	\end{subfigure}\hspace*{\fill}
	\begin{subfigure}[t]{0.1\textwidth}
		\centering
		\includegraphics[width=\linewidth]{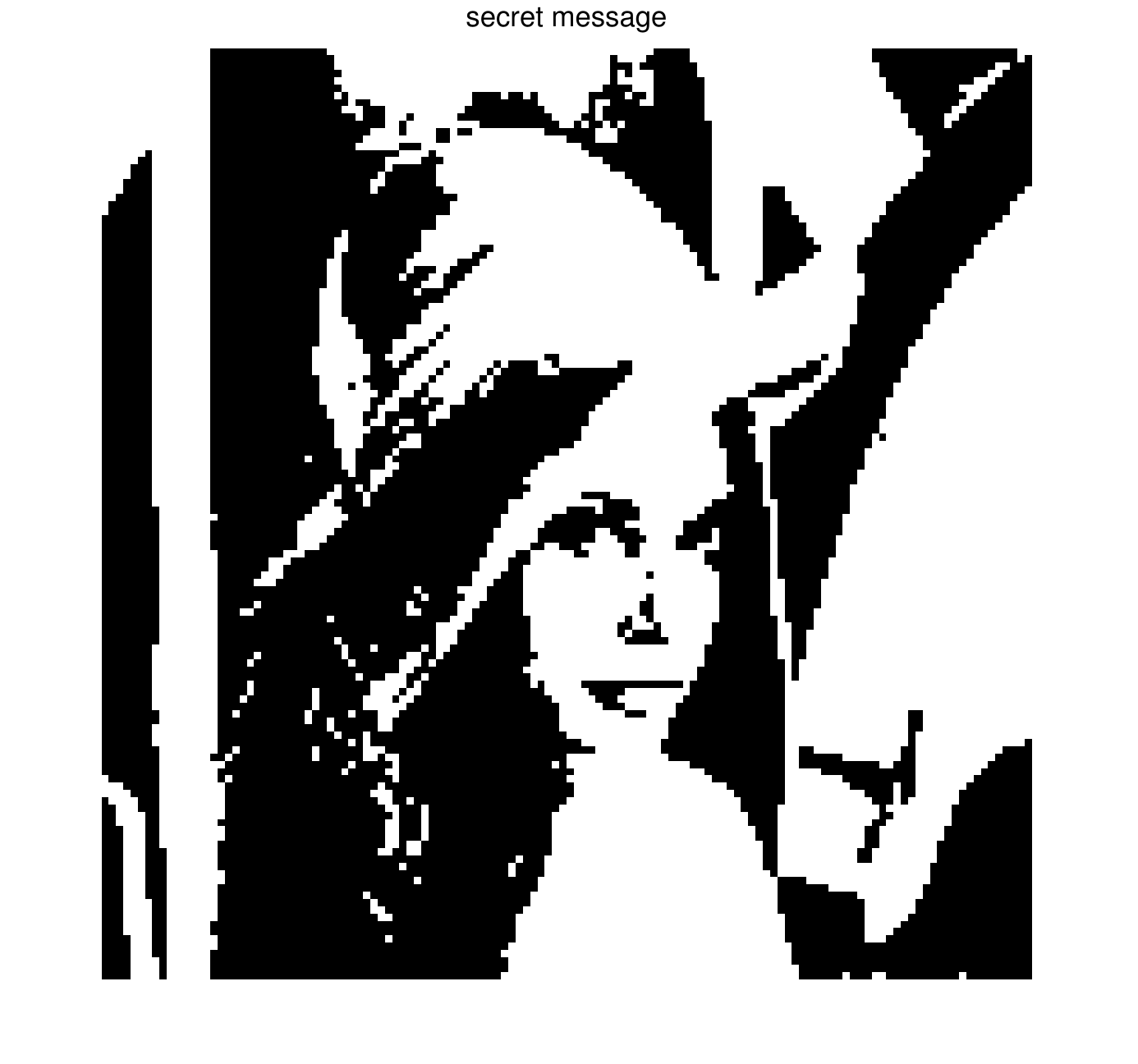}
		\caption{}
		\label{fig:side:a}
	\end{subfigure}\hspace*{\fill}
	\begin{subfigure}[t]{0.2\textwidth}
		\centering
		\includegraphics[width=\linewidth]{cover_256}
		\caption{}
		\label{fig:side:a}
	\end{subfigure}\hspace*{\fill}
	\begin{subfigure}[t]{0.2\textwidth}
		\centering
		\includegraphics[width=\linewidth]{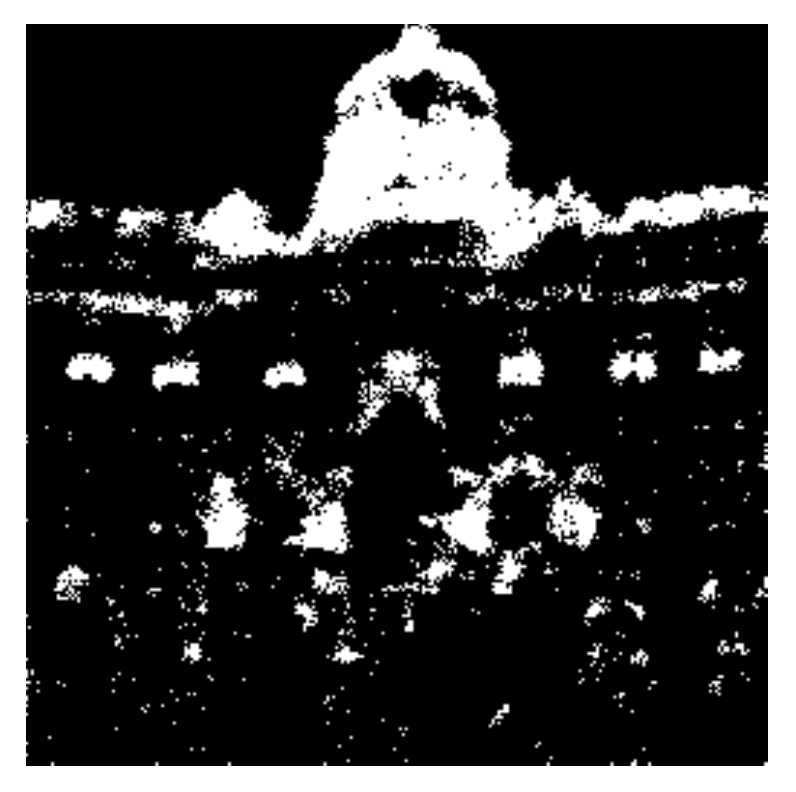}
		\caption{}
		\label{fig:side:a}
	\end{subfigure}\hspace*{\fill}
	\begin{subfigure}[t]{0.1\textwidth}
		\centering
		\includegraphics[width=\linewidth]{msg_lena}
		\caption{}
		\label{fig:side:a}
	\end{subfigure}\hspace*{\fill}
	\begin{subfigure}[t]{0.2\textwidth}
		\centering
	\end{subfigure}
	\caption{Illustration of the practical application of proposed UT-SCA-GAN. (a) The BOSSBase cover image ``1013.pgm" with a size of 256 $\times$ 256, (b) secrete message (with a size of 128 $\times$ 128), (c) stego image, (d) modification map, (e) recovered message.} \label{fig:Application}
\end{figure*}

We embed the binary image that has only two possible values 0 and 1 into the cover image. The flowchart of embedding and  extracting are show on Fig. \ref{fig:STC flowchart}. The embedding change probability denotes the probability learned form adversarial training. Fig. \ref{fig:Application} show an example of embedding process. Experimental results show that all of the secret message can be recovered by STC scheme, and the embedding positions are also located in complex region. Thus the proposed steganography scheme can improve the security performance and also can be used for practical application.

\section{Conclusion}
In this paper, a secure steganographic scheme based on generative adversarial network is proposed. A Tanh-simulator function was proposed to fit the optimal embedding simulator. A compact generator based on U-Net architecture is employed as the generator. To resist the current advanced steganalysis methods maxSRMd2, selection channel awareness (SCA) are incorporated into the discriminator. Experimental results showed that the proposed architecture outperforms the ASDL-GAN method dramatically by using less training time. Further more, it also obtained better performance than the hand-crafted method S-UNIWARD.

In our future work, we will explore the relationship between the architectures of generator and discriminator so as to boost the security performance. In addition, we would like to apply the proposed scheme to the JPEG domain.

\ifCLASSOPTIONcaptionsoff
\fi
\balance
\bibliographystyle{unsrt}
\bibliography{stegnography}

\end{document}